\documentstyle[preprint,aps,eqsecnum]{revtex}
\begin{document}
\draft
\tightenlines

\title{Multiphoton Radiative Recombination of Electron 
\\
Assisted by Laser Field}

\author{M.~Yu.~Kuchiev and V.~N.~Ostrovsky \cite{SP}
}

\address{School of Physics, University of New South Wales
Sydney 2052, Australia}

\maketitle

\begin{abstract}
In the presence of an intensive laser field the radiative 
recombination of the continuum electron into an
atomic bound state generally is accompanied 
by absorption or emission of several laser quanta.
The spectrum of emitted photons represents
an equidistant pattern with the spacing equal to the
laser frequency. The distribution of intensities in this 
spectrum is studied employing the Keldysh-type approximation, 
i.e. neglecting interaction of the impact electron with the atomic 
core in the initial continuum state. Within the adiabatic 
approximation the scale of emitted photon frequencies is 
subdivided into classically allowed and classically forbidden 
domains. The highest intensities correspond to emission 
frequencies close to the edges of classically allowed domain.
The total cross section of electron recombination summed
over all emitted photon channels exhibits negligible
dependence on the laser field intensity.
\end{abstract}



\section{INTRODUCTION} \label{Int}

It is well known that the laser plasma emits photons with frequencies
which are different from the frequency of the incident laser beam. 
For a number of applications an emission
of high energy photons is the most interesting phenomenon. 
The known  mechanisms which can be responsible 
for the high energy photo-production
could be identified as
the following three ones:
the high harmonic generation, laser stimulated bremsstrahlung and 
laser assisted recombination. 
These processes differ by the initial
and final states of the active electron. In the harmonic generation
the  initial state electron  occupies 
a (laser-dressed)  atomic bound state, usually it is the ground
atomic state.
In the  final  
state of this reaction 
the electron can occupy either the same bound state, or some 
excited or even ionized state.
The laser stimulated bremsstrahlung is a free-free transition
during which the electron is scattered by an atom in
the laser field.
During the 
scattering the electron emits a high-energy quantum.
In the laser assisted recombination (LAR) the electron starts in
the laser-dressed continuum, but ends up in the bound state.
The process of harmonic generation is currently studied very
actively with important advancements both in theory and experiment
(for reviews see Refs.~\cite{Protrev,Strelkov}). The laser
stimulated bremsstrahlung 
plays the very important role in plasma physics, see
recent experimental ~\cite{Ueshima,Norreys}
and theoretical ~\cite{Silin,Astapenko} works.

The subject of the present study is the LAR process.  As far as we know,
it has not yet received a proper attention in the literature, 
although its importance
for kinetics of laser plasma and its emission spectrum was
indicated before \cite{Zar}. 
From the point of view of the high-energy photo-production
the LAR possesses an advantage over the stimulated bremsstrahlung because
in LAR the electron impact energy is 
totally transferred to the high energy quanta.

The conventional (laser-field free) radiative recombination
of the continuum electron to the bound
state is a well studied process which is inverse to the photoionization. 
The frequency of the emitted photon  is uniquely defined by 
the energy conservation law. When a similar process occurs 
in the presence of an intensive laser field the radiation spectrum 
becomes 
much more richer since the recombination may be accompanied 
by absorption or emission of laser quanta. Therefore
the emitted photon spectrum  represents a sequence of 
equidistant lines separated by the laser frequency $\omega$. 
The recent review by Hahn \cite{Hahn} on the electron
recombination mentions only one, very special version
of LAR process, namely {\it one--photon}\/ 
LAR when the laser is tuned in resonance with
the energy of free-bound electron transition, and only 
emission of photons with this particular energy is considered.
The study of this special case was initiated quite long ago 
\cite{pos,Faisal,Neumann,Ritchie,Fill} and remains 
active in connection with the processes in the storage rings 
\cite{exp1,exp2,hypa,Na,Nb,exp,hypb}, formation of positronium 
\cite{pos} and antihydrogen \cite{ant,hypc,China} and even with 
possible cosmological manifestations \cite{cos}. In all these theoretical
studies the laser field was presumed to be weak and its influence 
on an initial and/or final electron states was neglected, except
Refs. \cite{Zar,China} which are commented below.

For production of high-energy photons
it is very interesting to extend the mentioned above studies allowing
for the multiphoton absorption during LAR.
Obviously the multiphoton processes 
can happen with high probability only in a strong laser field.
From this point of view there arises a necessary 
to fulfill a systematic study of LAR in a strong laser field in multiphoton
regime. This paper makes a first step in this direction. 

An additional, and rather unexpected
inspiration for the present study arises from
the fact that LAR 
comprises one of the steps in the three-step quantum scheme 
of high harmonic generation.
This scheme has recently been firmly established,
see Ref.\cite{HGG} and bibliography therein.
The major statement of \cite{HGG} is that
the high harmonic generation can be described as 
the multiphoton ionization of an atomic electron
which is followed by the LAR of this electron 
with the parent atomic particle.
From this point of view the LAR  plays a role of 
'a part' of the problem of  the high harmonic generation, 
which is important not only for the dense laser plasma,
but also for photo-production from individual atoms in strong laser fields,
where the harmonic generation 
is the major source for high energy photons.

The present study is devoted mostly 
to the patterns of intensities in the emitted photon spectrum 
depending on the laser field strength. We comment also on
the influence of laser field on the total recombination cross 
section. Here, as well as in other applications, the laser
field is intensive and LAR proceeds in substantially 
{\it multiphoton}\/ regime. 

Consider an electron in the laser-dressed continuum state 
$\Phi_{\bf p}(t)$ with the translational momentum ${\bf p}$.
Its recombination to the bound state generally results in
the emission of photons with the frequencies 
$\tilde{\Omega}_M$ defined from 
\begin{eqnarray} \label{phfr}
\tilde{\Omega}_M = \frac{1}{2} \, {\bf p}^2 + 
\frac{F^2}{4 \omega^2} - \varepsilon_a + M \omega ~,
\end{eqnarray}
where $\varepsilon_a$ is the quasienergy of the field-dressed
bound state $\Phi_a(t)$, ${\bf F}$ is the amplitude of the electric 
field strength in the laser wave, $F^2/(4 \omega^2)$ is 
the electron quiver energy in the laser field, $M$ is an integer.
Hereafter we use atomic system of units unless stated otherwise.
In the zero-laser-field limit ($F \rightarrow 0$) only emission 
of the photon with the frequency 
\begin{eqnarray} \label{Omegaff}
\Omega_{F \rightarrow 0} = \frac{1}{2} \, {\bf p}^2 + \left| E_a \right|
\end{eqnarray}
is allowed with $E_a$ being the bound state energy.
The presence of an intensive laser field makes possible multiphoton
processes when laser quanta are absorbed from the field 
or transmitted to it, with the amplitude 
\begin{eqnarray} \label{rec}
C_{M}({\bf p}) = \frac{1}{T}\int \limits_{0}^{T} dt \:
\langle \Phi_a (t) \mid \exp ( i \tilde{\Omega}_M t ) \, 
\hat{d}_{\mbox {\boldmath $\epsilon$}} \mid
\Phi_{\bf p} (t) \rangle ~, 
\quad \quad \quad
\hat{d}_{\mbox {\boldmath $\epsilon$}} = 
{\mbox {\boldmath $\epsilon$}} \cdot {\bf r} ~,
\end{eqnarray}
where $T = 2 \pi / \omega$ is the laser field period, and in 
the dipole momentum operator $\hat{d}_{\mbox {\boldmath $\epsilon$}}$
the unit vector ${\mbox {\boldmath $\epsilon$}}$
selects polarization of emitted radiation.
The LAR cross section is
\begin{eqnarray} \label{cr}
\sigma_M({\bf p}) = \frac{4}{3p} \: 
\frac{ \left( \tilde{\Omega}_M \right)^3}{c^3}
\: \left| C_M({\bf p}) \right|^2 ~, 
\end{eqnarray}
where $c$ is the velocity of light.
The cross section (\ref{cr}) refers to the process of 
{\it spontaneous}\/ LAR, since it is presumed that 
{\it incident}\/ electromagnetic field with the frequency
$\tilde{\Omega}_M$ is absent. In case if such a probe field 
is present, generally it would be amplified in course of
propagation through the medium containing free electrons. 
There is a number of theoretical works devoted to calculation 
of related gain in case of one-photon LAR. A recent paper by
Zaretskii and Nersesov \cite{Zar} explores the amplification 
in case of multiphoton LAR. Generally these studies imply 
some assumptions regarding the medium properties and result
in the expressions for the rate of stimulated transitions via 
that of spontaneous transitions and some characteristics of
laser beam and the experimental arrangement 
\cite{Faisal,Neumann,Ritchie,Zar}. The present paper 
provides analysis of spontaneous LAR whereas issues of radiation 
amplification are beyond its scope.

\section{KELDYSH-TYPE APPROXIMATION}

We develop the Keldysh-type approximation 
where the interaction of the continuum electron with the atomic core 
is neglected, i.e. the laser-dressed electron continuum state 
$\Phi_{\bf p}$ is approximated by the well-known Volkov state. 
The laser wave is assumed to be linear polarized with the
electric field strength ${\bf F}(t) = {\bf F} \cos \omega t$.
Explicit expression for the Volkov functions is conveniently cast as 
\begin{eqnarray}\label{Volkov}
\Phi_{\bf p}({\bf r},t) & = &
\chi_{\bf p}({\bf r},t) \, 
\exp \left( -i \bar E_{\bf p}t \right) ~,
\\ \label{chi}
\chi_{\bf p}({\bf r},t) & = &
\exp{\left \{ i \left [ ( {\bf p}+{\bf k}_t){\bf r} -
\int_{0}^{t} \left ( E_{{\bf p}}(\tau)-
\bar E_{\bf p} \right ) 
d \tau + \frac{{\bf p F} }{\omega^2} \right ] \right \} } ~,
\end{eqnarray}
where the factor $\chi_{\bf p}({\bf r},t)$ is time-periodic with
the period $T$, 
\begin{eqnarray}
{\bf k}_t & = & \frac{{\bf F}}{\omega} \sin \omega t ~,
\\ \label{insten}
E_{\bf p}(t) & = & 
\frac{1}{2} \left( {\bf p} + {\bf k}_t \right)^2 ~,
\\
\bar{E}_{\bf p} & = & 
\frac{1}{T} \, \int_0^T E_{\bf p}(\tau) \, d \tau =
\frac{1}{2} p^2 + \frac{F^2}{4 \omega^2} ~.
\end{eqnarray}
For the final bound state the field-free expression is employed 
\begin{eqnarray}\label{Psia}
\Phi_a({\bf r}, t) & = & \varphi_a({\bf r}) \, \exp( - i E_a t) ~,
\\ 
H_a \, \varphi_a({\bf r}) & = & E_a \, \varphi_a({\bf r}) ~,
\end{eqnarray}
where $H_a$ is the effective atomic Hamiltonian in the single 
active electron approximation.
The final bound state (\ref{Psia}) is always available if
electron collides with a positive ion. In case of collision with 
a neutral atom we assume existence of a stable negative ion.
By substituting formulae (\ref{Volkov})--(\ref{Psia})
into (\ref{rec}) one can see that the integrand is a periodic 
function of time provided the emitted photon frequency 
$\tilde{\Omega}_M$
satisfies (\ref{phfr}) with integer $M$ and $\varepsilon_a$
substituted by $E_a$. The lowest possible frequency of 
the {\it emitted}\/ photon is $\eta \omega$, where
\begin{eqnarray}
\eta = \frac{1}{\omega} \left( \frac{1}{2} p^2 
- E_a + \frac{F^2}{4 \omega^2} \right) -
{\rm Ent}\left[ \frac{1}{\omega} \left( \frac{1}{2} p^2 
- E_a + \frac{F^2}{4 \omega^2} \right)
\right] 
\end{eqnarray}
with ${\rm Ent}(x)$ being an integer part of $x$ ($0 \leq \eta < 1$). 
In the subsequent development we redefine labeling of emitted 
photon channels and instead of $\tilde{\Omega}_M$ (\ref{phfr}) 
employ the notation 
\begin{eqnarray}
\Omega_m = (m + \eta) \omega 
\quad \quad \quad (m \geq 0) ~.
\end{eqnarray}
The new label $m$ differs from the old one $M$ by an additive
integer. We find the labeling by $m$ more convenient
since it is rigidly related to the low-frequency edge of
the emitted photon spectrum: $m=0$ corresponds to the
lowest photon frequency $\eta \omega$.

By using the Fourier transformation formula (\ref{rec}) is 
rewritten as
\begin{eqnarray} \label{CF} 
C_{m}({\bf p}) = - \,\frac{1}{T} \, \int \limits_{0}^{T} dt \:
\exp \left\{ i \left[ (m + \eta) \omega t - S(t) \right] \right\} \:
\tilde{\varphi}_a^{(\epsilon)} \left( - {\bf p} - {\bf k}_t \right) ~,
\end{eqnarray}
where $S(t)$ is the classical action
\begin{eqnarray}
S(t) = \frac{1}{2} \int^t d \tau \left({\bf p} + {\bf k}_{\tau} \right)^2
-E_a t ~.
\end {eqnarray}
The function $\tilde{\varphi}_a^{(\epsilon)}({\bf q})$ is defined as
\begin{eqnarray} 
\tilde{\varphi}_a^{(\epsilon)}({\bf q}) = i 
\left( {\mbox {\boldmath $\epsilon$}} 
\cdot \nabla_{\bf q} \right) \tilde{\varphi}_a({\bf q}) ~. 
\end{eqnarray}
where $\tilde{\varphi}_a({\bf q})$ is the Fourier transform of 
the bound state wave function $\phi_a({\bf r})$:  
\begin{eqnarray} \label{wF}
\tilde{\varphi}_a({\bf q}) = \int d^3 {\bf r} \, 
\exp( - i {\bf q} {\bf r} ) \, \phi_a({\bf r}) ~ .
\end {eqnarray}

For the bound state wave function we use an asymptotic expression
\begin{eqnarray} \label{wf}
\phi_a({\bf r}) \approx A_a r^{\nu-1} \, \exp(- \kappa r) \,
Y_{lm}(\hat{{\bf r}})
\quad \quad \quad
( r \gg 1/\kappa),
\end{eqnarray}
where $\kappa = \sqrt{ 2 |E_a|}$, $\nu = Z/\kappa$, $Z$
is the charge of the atomic residual core ($\nu=Z=0$ for a
negative ion), $l$ is the active electron orbital momentum in
the initial state and $\hat{{\bf r}}$ is the unit vector.
The coefficients $A_a$ are tabulated for many negative ions \cite{RS}.
The Fourier transform $\tilde{\varphi}_a({\bf q})$ (\ref{wF})
is singular at $q^2 = \kappa^2$ with the asymptotic behavior for 
$q \rightarrow \pm i \kappa$ defined by the long-range asymptote
(\ref{wf}) in the coordinate space
\begin{eqnarray} \label{wfmo}
\tilde{\varphi}_a({\bf q}) = 4 \pi A_a (\pm 1)^l \, Y_{lm}(\hat{{\bf q}}) \,
\frac{(2 \kappa)^\nu \, \Gamma(\nu + 1)}{(q^2 + \kappa^2)^{\nu +1}} ~, 
\end{eqnarray}
where $(\pm 1)^l$ corresponds to $q \rightarrow \pm i \kappa$. 
In particular, for a negative ion ($\nu = 0$) with the active 
electron in an $s$ state ($l=0$) we have from (\ref{wfmo}) 
\begin{eqnarray}
\tilde{\varphi}_a({\bf q}) & = & \sqrt{4 \pi} A_a \: 
\frac{1}{(q^2 + \kappa^2)} ~, 
\\
\tilde{\varphi}_a^{(\epsilon)}({\bf q}) & = & - i  
\left( {\mbox {\boldmath $\epsilon$}} \cdot \hat{{\bf q}} \right)
\sqrt{4 \pi} A_a \: \frac{2 q}{(q^2 + \kappa^2)^2}  
\end{eqnarray}
($ \hat{{\bf q}} \equiv {\bf q}/q$ is unit vector).

\section{ADIABATIC APPROACH TO STIMULATED RECOMBINATION}

The time integral in (\ref{CF}) can be evaluated using 
the saddle point method. This amounts to the adiabatic 
approximation when the phase $(m + \eta) \omega t - S(t)$ 
in (\ref{CF}) is assumed to be large.  
The position of saddle points in the complex $t$-plane is 
governed by equation 
\begin{eqnarray} \label{spt}
S^\prime(t_{m \mu}) - \Omega_m = 0 ~,
\end{eqnarray}  
or, more explicitly,
\begin{eqnarray} \label{sptexp1}
\frac{1}{2} \left({\bf p} + {\bf k}_{t_{m \mu}} \right)^2 = 
E_a + (m + \eta) \omega ~.
\end{eqnarray}  
It is convenient to single out in the electron momentum vector
${\bf p} = {\bf p}_\parallel + {\bf p}_\perp$ components parallel 
(${\bf p}_\parallel$) and perpendicular (${\bf p}_\perp$)
to the electric field vector ${\bf F}$. Then Eq.(\ref{sptexp1})
is rewritten as 
\begin{eqnarray} \label{sptexp}
\frac{1}{2} \left(p_\parallel + k_{t_{m \mu}} \right)^2 = 
E_a - \frac{1}{2} p^2_\perp + (m + \eta) \omega ~.
\end{eqnarray} 
For each value of $m$ this equation has a number of solutions 
$t_{m \mu}$ distinguished by the extra subscripts $\mu$.
In the saddle point approximation the time integration in 
formula (\ref{rec}) is cast as
\begin{eqnarray} 
\label{Csp}
C_{m}({\bf p}) & = & - \,
\frac{1}{T} \: \sum_\mu \: 
\sqrt{\frac{2 \pi}{ i S^{\prime \prime} \left(t_{m \mu} \right)}} \:
\exp \left\{ i \left[\Omega_m t_{m \mu} - S(t_{m \mu}) \right] \right\} 
\:
\tilde{\varphi}_a^{(\epsilon)} \left( - {\bf p} - {\bf k}_{t_{m \mu}} 
\right) ~,
\end{eqnarray}
where summation is to be taken over the saddle points $t_{m \mu}$
operative in the contour integration 
$\left[{\bf k}_{t_{m \mu}} = ({\bf F}/\omega) \sin \omega t_{m \mu}
\right]$).

The saddle points are found from Eq.(\ref{sptexp}) as
\begin{eqnarray} \label{sprad}
\sin \omega t_{m \mu} = \frac{\omega}{F} 
\left(- p_\parallel \pm \sqrt{2 (m + \eta) \omega - \kappa^2 
- p^2_\perp} \right) ~.
\end {eqnarray}
The subscript $\mu$ labels solutions differing by the choice of 
the sign in (\ref{sprad}) and sign in 
$\cos \omega t_{m \, \mu} = \pm \sqrt{1 - \sin^2 \omega t_{m \mu}}$.
There are four solutions per the laser field cycle 
(i.e for $0 \leq {\rm Re} \, t_{m \mu} < T$). 

In order to elucidate the meaning of the saddle point equation 
(\ref{sptexp}) we rewrite it as
\begin{eqnarray}
E_{\rm p}(t_{m \mu}) - E_a = \Omega_m ~.
\end{eqnarray}
It shows that the photons are preferentially emitted at 
the moment of time
when instantaneous continuum electron energy $E_{\rm p}(t)$ 
(\ref{insten}) is separated from the bound state energy $E_a$ 
by the energy of the emitted photon $(m + \eta) \omega$. 
The LAR process is most effective when this occurs at some real
moment of time, i.e. the saddle points $t_{m \mu}$ are real-valued. 
This regime corresponds to the {\it classically allowed radiation}.
It can happen only for some part of the emitted photon spectrum,
i.e. only in some domain of $m$. Outside it, when $t_{m \mu}$
possesses an imaginary part, the emission is 
strongly suppressed. Remarkably, within the classically allowed
domain the intensity of emitted lines could vary very significantly
as detailed below.

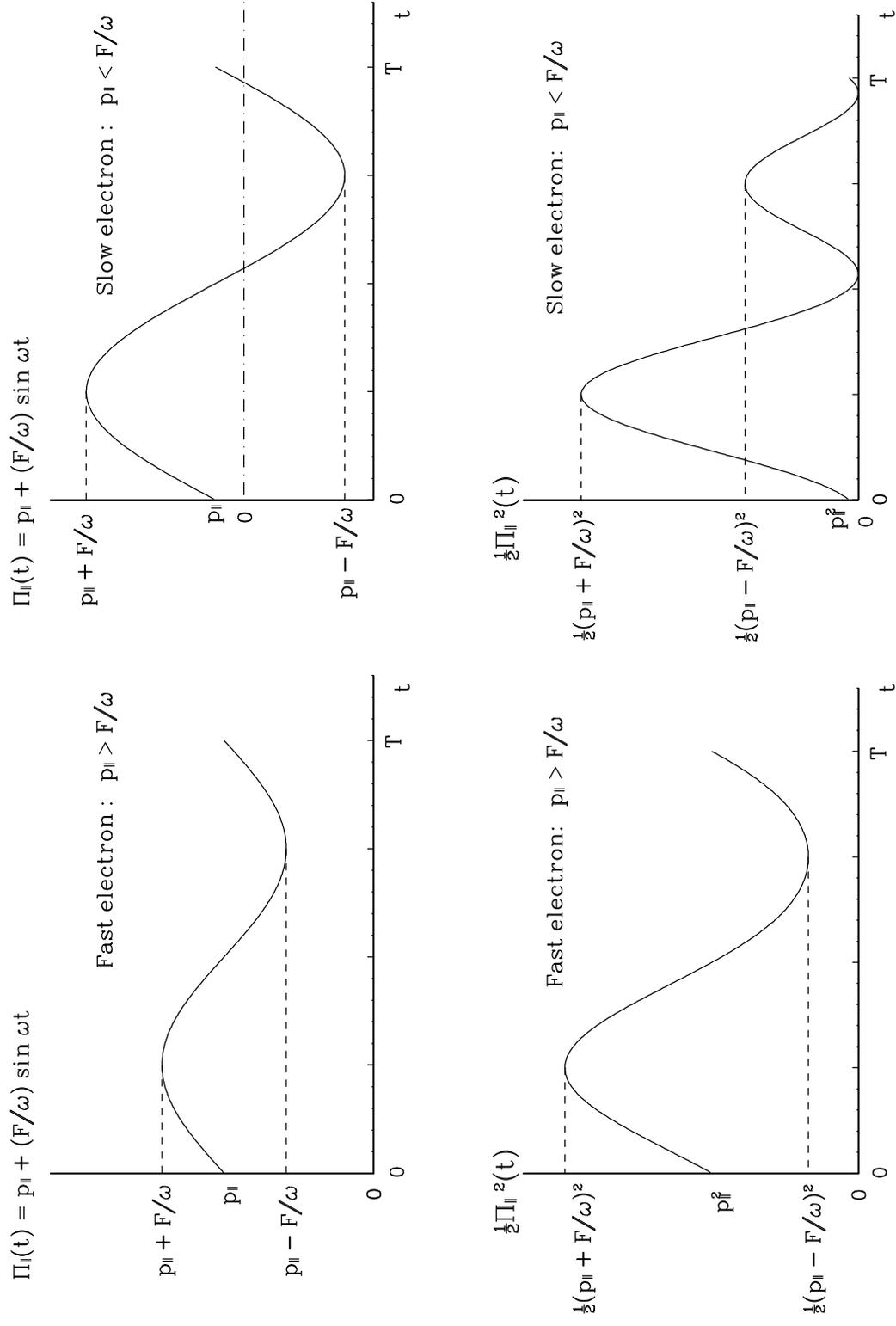
\begin{figure}
\caption{ \label{scheme}
Regimes of fast ($p_\parallel > F/\omega$) and 
slow ($p_\parallel < F/\omega$) 
electron in the laser-assisted recombination process. 
For each regime the schematic plots show electron momentum
with account for the quiver motion in laser field
$\Pi(t)_\parallel \equiv p_\parallel + (F/\omega) \sin \omega t$
and the effective instantaneous kinetic energy 
$\frac{1}{2} \Pi(t)_\parallel$. As time $t$ varies, the function
$\frac{1}{2} \Pi(t)_\parallel$ oscillates in the interval that 
covers the emitted photon energies $\Omega_m$ allowed for 
population classically. Outside this interval only 
non-classical (tunneling) population is possible. 
Fig. \protect\ref{scheme}a shows that in the classically 
allowed domain each value of the photon energy 
$\Omega$ is passed twice during the laser field period $T$ 
if the electron is fast ($p_\parallel > F/\omega$). 
In the slow electron regime ($p_\parallel < F/\omega$) 
the classically allowed domain of $\Omega_m$ is subdivided 
into two regions, as seen from Fig. \protect\ref{scheme}.
The photons with higher $\Omega_m$ are again emitted in 
the double-passage mode, whereas the lower values of 
$\Omega_m$ are passed four times per the laser field cycle.
}
\end{figure}

The necessary condition of classically allowed radiation,
\begin{eqnarray}
\Omega_m > \left| E_a \right| + \frac{1}{2} p^2_\perp ~,
\end{eqnarray}
makes real the right hand side of formula (\ref{sprad}). 
Details of classically allowed emission 
depend on the relation between the electron translational
momentum component $p_\parallel$ and the momentum $F/\omega$ 
acquired by the electron in its quiver motion in the laser field.
In the fast electron regime, $p_\parallel > F/\omega$, the term 
$\frac{1}{2}(p_\parallel + k_t)^2$ never passes zero as time $t$
varies. 
As a result, the saddle point equation (\ref{sptexp1}) 
has two or zero real-valued solutions per field cycle 
(in the classically allowed and forbidden domains respectively,
see Fig. \ref{scheme}a).
In the slow electron case, $p_\parallel < F/\omega$, the 
$\frac{1}{2}(p_\parallel + k_t)^2$ passes via zero.
Due to this circumstance, as seen from Fig. \ref{scheme}b, 
for some interval of photon frequencies $\Omega_m$ 
the equation (\ref{sptexp1}) has four real-valued solutions
whereas for higher $\Omega_m$ only two solutions exist.
Consequently, in this case the classically allowed
domain is subdivided in two parts. The related LAR regimes
are discussed below in more detail.

\subsection{Fast electron regime: $p_\parallel > F/\omega$}

Here one has to choose the upper sign in formula (\ref{sprad})
in order to get a real-valued saddle point. The condition 
$\left| \sin \omega t_{m \mu} \right| \leq 1$
is straightforwardly reduced to
\begin{eqnarray} \label{cond1}
\frac{1}{2} \left( p_\parallel - \frac{F}{\omega} \right)^2 
+ \left| E_a \right| + \frac{1}{2} p^2_\perp \leq \Omega_m \leq
\frac{1}{2} \left( p_\parallel + \frac{F}{\omega} \right)^2 
+ \left| E_a \right| + \frac{1}{2} p^2_\perp ~.
\end{eqnarray}
In this photon frequency interval only one pair of real 
saddle points $t_{m \mu}$ exists per field cycle, see Fig \ref{scheme}. 
These two saddle points are to be included
into summation over $\mu$ in (\ref{Csp}). The phase difference
between the two terms in (\ref{Csp}) varies with $m$.
As a result $\left|C_m({\bf p})\right|^2$ oscillates 
between zero and some envelope function $\Xi(m)$ defined as
\begin{eqnarray} \label{env}
\Xi(m) = \frac{8 \pi}{T^2 S^{\prime \prime}}
\left| \tilde{\varphi}_a^{(\epsilon)} 
\left( - {\bf p} - {\bf k}_{t_{m \mu}} \right) \right|^2 ~,
\\
\left| \tilde{\varphi}_a^{(\epsilon)} 
\left( - {\bf p} - {\bf k}_{t_{m \mu}} \right) \right|^2 
= \pi A_a^2 \: \frac{2(m+\nu) \omega - \kappa^2 - p^2_\perp}
{(m + \nu)^4 \omega^4} ~,
\\
S^{\prime \prime} = F \sqrt{2(m+\nu) \omega - \kappa^2 - p^2_\perp} \: \:
\sqrt{ 1 - \frac{\omega^2}{F^2} \left(p_\parallel - 
\sqrt{2(m+\nu) \omega - \kappa^2 - p^2_\perp} \right)^2} ~.
\end{eqnarray}
As could be anticipated, the function $\Xi(m)$ has weak singularities
at the boundaries of the classically allowed region.
The extension of the classically allowed region on the photon
frequency scale is $2 p_\parallel F /\omega$ with its center located
at $\Omega_{\rm c} = \frac{1}{2} p^2 + \left| E_a \right|
+ F^2 / (2 \omega^2)$. For vanishing laser field
$\Omega_{\rm c}$ tends to the limit (\ref{Omegaff})
and the classically allowed domain shrinks to the single line. 
The condition that a single line dominates in the photon spectrum 
could be formulated as $2 p_\parallel F / \omega^2 \sim 1$.

Fig.~\ref{Fig1} illustrates evolution of the spectrum
pattern with the laser intensity $I$.
We consider electrons with the energy 
$E_{\rm el} = \frac{1}{2} p^2$ equal to 1 eV ($p=0.271$) 
in the laser field with the frequency $\omega = 0.0043$
and different intensities.
The electron momentum ${\bf p}$ is directed along the
laser field strength ${\bf F}$ ($p_\perp = 0$). 
The electron recombines to the bound state of H$^-$ ion 
($\kappa = 0.2354$, $A_a = 0.75$ \cite{RS}).
The emission amplitudes are obtained by numerical evaluation of 
the time integral in (\ref{CF}). The laser field intensities 
$I = 10^{11}, \: 10^{10}, \: 10^{9}, \: 10^{8}, \: 10^{7}$ W/cm$^2$
corresponds to the values of parameter $2 p F / \omega^2$
respectively $49.4, \: 15.6, \: 4.94, \: 1.56, \: 0.494$.
For the weakest field considered ($I = 10^{7}$ W/cm$^2$)
the intensity of the principal line in the spectrum ($m=14$) 
exceeds more than 50 times these of adjacent satellites.
For $I = 10^{8}$ W/cm$^2$ this ratio is substantially smaller 
($\sim 5$). When laser field is increased by an order 
of magnitude, the dip in the emitted photon spectrum 
appears at $m=15$. This is the first manifestation of the 
oscillatory structure in the spectrum due to interference 
of two contributions in (\ref{Csp}). For $I = 10^{10}$ W/cm$^2$ 
the structure becomes well manifested.
At last, for $I = 10^{11}$ W/cm$^2$ the structure becomes 
well developed and extended. In the latter case, in fact,
the situation is beyond the fast electron regime; 
it will be discussed in the next subsection.

The semiclassical formula (\ref{Csp}) is applicable when the 
classically allowed domain is sufficiently broad on the 
frequency scale. Fig. \ref{Fig2} shows the photon spectrum 
in the well manifested semiclassical regime ($E_{\rm el} = 10$ eV, 
$I = 10^{11}$ W/cm$^2$, $p = 0.857$, $F/\omega = 0.392$). 
In the classically allowed domain ($31 \leq m \leq 187$)
the quantities $\left|C_m(p) \right|^2$ obtained by 
numerical evaluation of the integral (\ref{CF})
over time (circles) oscillate violently  
due to the interference effects. Outside this region 
$\left| C_m(p) \right|^2$ decrease very rapidly. 
Note that the most efficient emission occurs at the edges
of the classically allowed  interval. This effect is completely
analogous to enhancement of the probability density near
the turning points for the quantum particle moving
in the potential well.
The envelope function (\ref{env}) (solid curve) reproduces 
well this overall behavior. The saddle point approximation
(\ref{Csp}) allows us to reproduce well the oscillatory 
structure (squares in Fig. \ref{Fig2}). Within the 
classically-allowed domain the summation in this formula runs
over two real-valued saddle points $t_{m \mu}$.
As $m$ varies approaching the domain border, two saddle 
points lying at the real-$t$ axis
approach each other and eventually merge at the boundary. 
After that they separate again moving perpendicular to the real
axis in the complex $t$-plane. The latter situation
corresponds to the classically forbidden, or tunneling
regime where only one saddle point is to be included
in the summation over $\mu$ in (\ref{sprad})
(namely, that which ensures exponential decrease
of $\left|C_m(p) \right|^2$ outside the classically-allowed 
domain). The transition between two regimes could be described
by the Airy function. We do not pursue here the detailed
description of this, rather standard situation. 
In particular, Fig.~\ref{Fig2}, the results shown by 
squares in Fig.~\ref{Fig2} are obtained using the plain
semiclassical formula (\ref{Csp}) with two or one
saddle points included as discussed above; the deviations from
the numerical results are seen to be essential only in a very 
narrow transitions region.
Since the numerical evaluation of integral (\ref{CF}) over 
time is not difficult, we employ the adiabatic approach in
order to obtain better insight into the pattern of
emitted radiation spectrum, but not for producing
an alternative method to evaluate the amplitudes.

\subsection{Slow electron regime: $p_\parallel < F/\omega$}

In this case the real-valued result for $t_{m \mu}$ is
provided by both upper and lower sign in the expression
(\ref{sprad}). It is easy to see from Fig. \ref{scheme}b
that the classically
allowed region of photon frequencies is subdivided in 
two domains. The first of them, with one pair of real-valued 
saddle points $t_{m \mu}$, corresponds to $\Omega_m$ lying
in the interval (\ref{cond1}). At smaller photon frequencies, 
another subdomain is defined by the condition
\begin{eqnarray} \label{cond2} 
\left| E_a \right| + \frac{1}{2} p^2_\perp \leq \Omega_m \leq
\frac{1}{2} \left( p_\parallel - \frac{F}{\omega} \right)^2 
+ \left| E_a \right| + \frac{1}{2} p^2_\perp ~.
\end{eqnarray}
Here {\it two pairs}\/ of real saddle points $t_{m \mu}$ exist.
The spectrum for this situation is illustrated by Fig. \ref{Fig3} 
($E_{\rm el} = 0.1$ eV, $I = 10^{11}$ W/cm$^2$, $p = 0.0857$, 
$F/\omega = 0.392$). 
The classically allowed domain lies in the interval 
$7 \leq m \leq 32$, with the four-saddle-point regime being operative 
for $7 \leq m \leq 17$, and the two-saddle point regime
for $18 \leq m \leq 32$. The results of numerical calculations
shown by circles suggest that the oscillations in
$\left| C_m(p) \right|^2$ or $\sigma_m(p)$ proceed with
two different frequencies, the higher frequency being characteristic 
for the four-saddle-point domain. The plain semiclassical formula 
(\ref{Csp}) (squares) essentially reproduces this structure. 
Of course, it is not designed for accurate description 
of a transition between the two-saddle-point and four-saddle-point 
regimes where the deviations are seen to be larger.
A special, more sophisticated treatment is required here, but 
such complications are not pursued in the present study as
argued above. The non-standard situation emerges also
at the left edge of the classically allowed interval where
all saddle points simultaneously move from the real axis into 
the complex $t$ plane. This transition region could not be 
described by a simple Airy-type pattern that is known to give 
a monotonous decrease in the classically forbidden domain; 
on the contrary, the numerical results reveal some structure 
in this region, see Fig. \ref{Fig3}. Bearing all this in mind
it is not unexpected that the plain semiclassical approximation 
(\ref{Csp}) essentially fails near the left border of the
classically allowed domain. 

It is worthwhile to mention also another region where the standard
semiclassical approximation fails. Namely, for $\Omega = 0$
the saddle point positions coincide with the poles of the function
$\tilde{\varphi}_a^{(\epsilon)}$. The situation when an exact
coincidence occurs is tractable rather easily \cite{GK}. 
Somewhat more effort is required to obtain uniform
description of a transition between this case and a situation
when the saddle point and the pole are well separated, as presumed
in simple formula (\ref{Csp}). Again, such sophistication 
are beyond the scope of the present study.

At last, Fig.~\ref{Fig4} shows a transient situation between 
the fast and slow electron regimes ($E_{\rm el} = 1$ eV, 
$I = 10^{11}$ W/cm$^2$, $p = 0.271$, $F/\omega = 0.392$).
Here only two harmonics ($ m = 7, \: 8$) correspond to the 
four-saddle-point regime. The remaining part of the
classically-allowed domain, $9 \leq m \leq 56$ corresponds
to two-saddle-point regime. Most of the spectrum is well 
described by the plain saddle-point approximation (\ref{Csp})
and covered by the envelope function (\ref{env}), albeit the 
highest peak at $m=9$ exceeds it, as being in the region
of the transition between the two and four-saddle point regimes.
Quite paradoxically, the low-frequency classically
forbidden region with well manifested structure
exhibits much higher emission intensities as compared with
the large-frequency edge of the classically allowed domain.

\section{CONCLUSION}

As discussed in the Introduction, the LAR is one of the processes
responsible for emission of high energy photons by the laser plasma.
Surprisingly, it has not yet received attention of researchers.
This is particularly unsatisfactory since the other processes leading
to high energy photons (harmonic generation and laser stimulated
bremsstrahlung) are currently under active scrutiny. The present 
paper could be considered as a first step to start filling this gap.  
The theory in many aspects is parallel to the treatment of 
multiphoton ionization (MPI) where the Keldysh approximation is known 
to provide an important insight and quantitatively reliable results.
The origin of differences between MPI and LAR lies 
in the kinematics: in MPI process the allowed electron 
energy in the continuum are robustly defined by the parameters
of the system (initial electron binding energy, laser field
frequency and strength), whereas in the LAR the continuum electron energy 
is arbitrary. This rather trivial observation results in 
important consequences of physical character. They are particularly 
lucid in the adiabatic regime when laser frequency is sufficiently 
small. The ionization is a {\it tunneling}\/ process for all 
above-threshold channels. On the contrary, in the LAR there 
is a domain of photon frequencies for which emission is allowed 
{\it classically}. 

The Keldysh-type approximation allowed us to
describe evolution of the LAR spectrum as the laser field varies,
from the single line with only weak satellites in the low-field 
limit to the broad pattern of equidistantly spaced harmonics 
in the strong field case. In the adiabatic approximation 
(i.e. the saddle point method) the photon spectrum is subdivided into 
classically allowed and classically forbidden domains, with 
the line intensities being highest at the boundaries  
of the former region. Concerning the quantitative side of the
problem, the adiabatic approach is less efficient
for the LAR process as compared with the treatment of above
threshold ionization (ATI). The reason is that in the latter case
the saddle point method is well applicable in its most simple form,
whereas for LAR process some technical complications emerge.
The difference stems from the fact that ATI process always
effectively occurs at complex-valued moments of time,
whereas for LAR this is generally not the case, and several
regimes could be operative with the transition regions between 
them. Albeit not drastic, these complications to our opinion 
hardly warrant necessary cumbersome analytical involvements, 
bearing in mind that the numerical calculations are quite simple 
and straightforward. Nevertheless the saddle point method
remains very useful for understanding the intensity 
patterns in the emitted photon spectrum.

An additional assumption of the present study, that in principle 
could be easily abandoned, is the use of asymptotic expression 
(\ref{wf}) for the final bound state 
wave function. Again, in the LAR process the situation is less 
favorable for this approximation as compared with 
the ATI process. This is because, as discussed in detail
earlier \cite{GK}, the long-range asymptote of the bound
state wave function governs ATI amplitudes, whereas
LAR process is more sensitive to the wave function behavior 
in the entire coordinate space.

As is pictured by Fig. \ref{Fig1}, the amount of 
noticeable lines in the photon spectrum increases with the 
laser field strength, but the intensity of each individual 
line decreases in average. The cross section of the electron 
transition into the bound state summed over all emitted
photon channels is 
$\sigma_{\rm tot}(p) = \sum_{m>0} \sigma_m(p)$. It exhibits
only very weak dependence on the laser filed intensity $I$
\cite{ftnt}. For instance, in the particular case of Fig.~\ref{Fig1}
we obtain for $\sigma_{\rm tot}(p)$ the values
$3.85 \cdot 10^{-6}$, $3.85 \cdot 10^{-6}$,
$3.89 \cdot 10^{-6}$, $3.64 \cdot 10^{-6}$,
$3.4 \cdot 10^{-6}$ for the laser field intensities
$I = 10^{11}, \: 10^{10}, \: 10^{9}, \: 10^{8}, \: 10^{7}$ W/cm$^2$.
Recent calculations \cite{China} of the laser-assisted
antihydrogen formation in positron-antiproton
collisions employed Coulomb-Volkov wave function for
the initial electron continuum state $\Phi_{\bf p}$
and the laser-perturbed wave function for the bound state.
The authors considered only one-photon LAR process and concluded
that the LAR cross section decreases for  the stronger laser fields.
The present results indicate that if the 
multiphoton processes are included, then the total LAR cross section
is essentially independent on laser field intensity.

Thus the effect of a laser on the recombination process looks
very straightforward. The total cross section of recombination 
essentially is not changed by a laser field, but is redistributed over 
equidistant pattern in photon spectrum that becomes broader
as the laser intensity increases.

\acknowledgements

This work has been supported by the Australian Research Council. 
V.~N.~O. acknowledges the hospitality of the staff of 
the School of Physics of UNSW where this work has been
carried out.

\begin{figure}
\caption{ \label{Fig1}
Factor $\left| C_m(p) \right|^2$ and cross section
$\sigma_m(p)$ for laser-assisted recombination of the electron 
with the energy $E_{\rm el} = 1$ eV to the bound state in H$^-$ ion.
The results of numerical integration in Eq.(\protect\ref{CF})
are shown for the laser field with the frequency 
$\omega = 0.0043$ and the intensities $I = 10^7$ W/cm$^2$ (crosses);
$10^8$ W/cm$^2$ (triangles); $10^9$ W/cm$^2$ (diamonds);
$10^{10}$ W/cm$^2$ (squares) and $10^{11}$ W/cm$^2$ (circles).
The symbols are joined by lines to help the eye.
}
\end{figure}

\begin{figure}
\caption{ \label{Fig2}
Same as in Fig. \protect\ref{Fig1}, but for the electron 
energy $E_{\rm el} = 10$ eV and the laser field intensity
$I = 10^{11}$ W/cm$^2$. The results of numerical calculations 
and plain semiclassical formula (\protect\ref{Csp})
are shown respectively by circles and squares. 
The semiclassical envelope function
(\protect\ref{env}) is given by solid line. 
In the zero-laser-field limit the spectrum shrinks to the
single line with the position indicated by vertical arrow.
}
\end{figure}

\begin{figure}
\caption{ \label{Fig3}
Same as in Fig. \protect\ref{Fig2}, but for the electron 
energy $E_{\rm el} = 0.1$ eV and the laser field intensity
$I = 10^{11}$ W/cm$^2$. 
}
\end{figure}

\begin{figure}
\caption{ \label{Fig4}
Same as in Fig. \protect\ref{Fig2}, but for the electron 
energy $E_{\rm el} = 1$ eV and the laser field intensity
$I = 10^{11}$ W/cm$^2$. 
}
\end{figure}


\begin{references} 
\bibitem[*]{SP} 
Permanent address:  
Institute of Physics, The University of St Petersburg, 
198904 St Petersburg, Russia;  E-mail: Valentin.Ostrovsky@pobox.spbu.ru


\bibitem{Protrev}
M.~Protopapas, C.~H.~Keitel, and P.~L.~Knight, Rep. Progr. Phys.
{\bf 60}, 389 (1997).

\bibitem{Strelkov}
V.~T.~Platonenko and V.~V.~Strelkov, 
Kvantovaya Elektronika {\bf 25}, 582 (1998)
[Quantum Electronics {\bf 28}, 584 (1998)].

\bibitem{Ueshima}
Y.~Ueshima,  Y.~Kishimoto,  A.~Sasaki,  T.~Tajima,
Laser Part. Beams. {\bf 17}, 45 (1999).

\bibitem{Norreys}
P.~A.~ Norreys  M.~Santala,  E.~Clark,  M.~Zepf,  I.~Watts.  
F.~N.~ Beg,  K.~Krushelnick,  
M.~Tatarakis,  A.~E.~Dangor,  X.~Fang,  P.~Graham,  T.~McCanny, R.~P.~Singhal, 
  K.~W.~D.~ Ledingham,  A.~Creswell,  D.~C.~W.~Sanderson,  J.~Magill,  
A.~Machacek,  J.~S.~Wark,  R.~Allott,  B.~Kennedy,  D.~Neely.
  Phys. Plasmas. {\bf 6}  2150 (1999).


\bibitem{Silin}
V.~P.~Silin,
  Izv. Akad. Nauk Ser. Fiz. {\bf 63}, 707 (1999).

\bibitem{Astapenko}
V.~A.~ Astapenko,  Laser Phys. {\bf 8}, 1066 (1998).



\bibitem{Zar}
D.~F.~Zaretskii and E.~A.~Nersesov, 
Zh. Eksp. Teor. Fiz. {\bf 109}, 1994 (1996)
[JETP {\bf 82}, 1073 (1996)].

 
\bibitem{Hahn}
Y.~Hahn, Rep. Progr. Phys. {\bf 60}, 691 (1997).

\bibitem{pos}
L.~A.~Rivlin, Kvantovaya Elektronika {\bf 6}, 594 (1979)
[Sov. J. Quant. Electron {\bf 9}, 353 (1979)].

\bibitem{Faisal}
F.~H.~M.~Faisal, A.~Lami, and N.K.Rahman, J. Phys. B {\bf 14}, 
L569 (1981);  A.~Lami, N.K.Rahman, and F.~H.~M.~Faisal, 
Phys. Rev. A {\bf 30}, 2433 (1984).

\bibitem{Neumann}
R.~Neumann, H.~Poth, A.~Winnacker, and A.~Wolf,
Z. Phys. A {\bf 313}, 253 (1983).

\bibitem{Ritchie}
B.~Ritchie, Phys. Rev. A {\bf 30}, 1849 (1984).

\bibitem{Fill}
E.~F.~Fill, Phys. Rev. Lett. {\bf 56}, 1687 (1986).

\bibitem{exp1}
U.~Schramm, J.~Berger, M.~Grieser, D.~Habs, E.~Jaeschke, G.~Kilgus,
D.~Schwalm, and A.~Wolf, Phys. Rev. Lett. {\bf 67}, 22 (1991).

\bibitem{exp2}
F.~B.~Yousif, P.~Van der Donk, Z.~Kucherovsky, J.~Reiss, E.~Brannen,
J.~B.~A.~Mitchell, and T.~J.~Morgan, Phys. Rev. Lett. 
{\bf 67}, 26 (1991).

\bibitem{hypa}
U.~Schramm, T.~Schl\"{u}ssler, D.~Habs, D.~Schwalm, and
A.~Wolf, Hyperfine Interactions {\bf 99}, 309 (1996).

\bibitem{Na}
S.~Pastuszka, U.~Schramm, M.~Grieser, C.~Broude, R.~Grimm,
D.~Habs, J.~Kenntner, H.-J.~Miesner, T.~Sch\"{u}ssler, 
D.~Schwalm, and A.~Wolf, Nucl. Inst. Meth. A {\bf 369}, 
11 (1996).

\bibitem{Nb}
S.~Asp, R.~Schuch, D.~R.~DeWitt, C.~Biedermann, H.~Gao,
W.~Zong, G.~Andler, E.~Justiniano, Nucl. Inst. Meth. B {\bf 117}, 
31 (1996).

\bibitem{exp}
M.~L.~Rogelstad, F.~B.~Yousif, T.~J.~Morgan, and J.~B.~A.~Mitchell,
J. Phys. B {\bf 30}, 3913 (1997)

\bibitem{hypb}
E.~Justiniano, G.Andler, S.~Asp, D.~R.~DeWitt, and R.Schuch,
Hyperfine Interactions {\bf 108}, 283 (1997).

\bibitem{ant}
R.~Neumann, H.~Poth, A.~Winnacker, and A.~Wolf,
Z. Phys. A {\bf 313}, 253 (1983).

\bibitem{hypc}
A.~M\"{u}ller and A.~Wolf, Hyperfine Interactions {\bf 109}, 233
(1997).

\bibitem{China}
S.-M.~Li, Y.-G.~Miao, Z.-F.~ Zhou, J.~Chen and Y.-Y.~Liu,
Phys. Rev. A {\bf 58}, 2615 (1998).

\bibitem{cos}
W.~Klemperer, X.-C. Luo, R.~Rosner, and D.~N.~Schramm,
Proc. Nat Ac. Sci. USA {\bf 92}, 6166 (1995).

\bibitem{HGG}
M.~Yu.~Kuchiev and V.~N.~Ostrovsky, J.Phys.B {\bf 32}, L189 (1999);
Phys. Rev. A (accepted for publication).

\bibitem{RS}
A.~A.~Radzig and B.~M.~Smirnov {\it Reference Data on Atoms,
Molecules and Ions} (Berlin: Springer, 1985).
Unfortunately the numerical value of the asymptotic parameter
$A$ for H$^-$ ion is absent in the standard reference book
[A.~A.~Radzig and B.~M.~Smirnov, {\it Reference Data on Atoms,
Molecules and Ions}\/ (Berlin: Springer, 1985)].
In our calculations, as previously \cite{GK} \cite{HGG}, 
we assume $A=0.75$ as given by V.~M.~Galitzkii, E.~E.~Nikitin, 
and B.~M.~Smirnov, {\it Teoriya Stolknovenii Atomnykh Chastitz}
(In Russian: {\it Theory of Atomic Particle Collisions})
(Moscow: Nauka, 1981). 

\bibitem{GK}
G.~F.~Gribakin and M.~Yu.~Kuchiev, Phys. Rev. A {\bf 55}, 3760 (1997);
J. Phys. B {\bf 30}, L657 (1997); {\bf 31}, 3087 (1998);
M.~Yu.~Kuchiev and V.~N.~Ostrovsky, J. Phys. B {\bf 31}, 2525 (1998).

\bibitem{ftnt}
Unfortunately currently we did not succeed in analytical derivation
of this result.

\end{references}
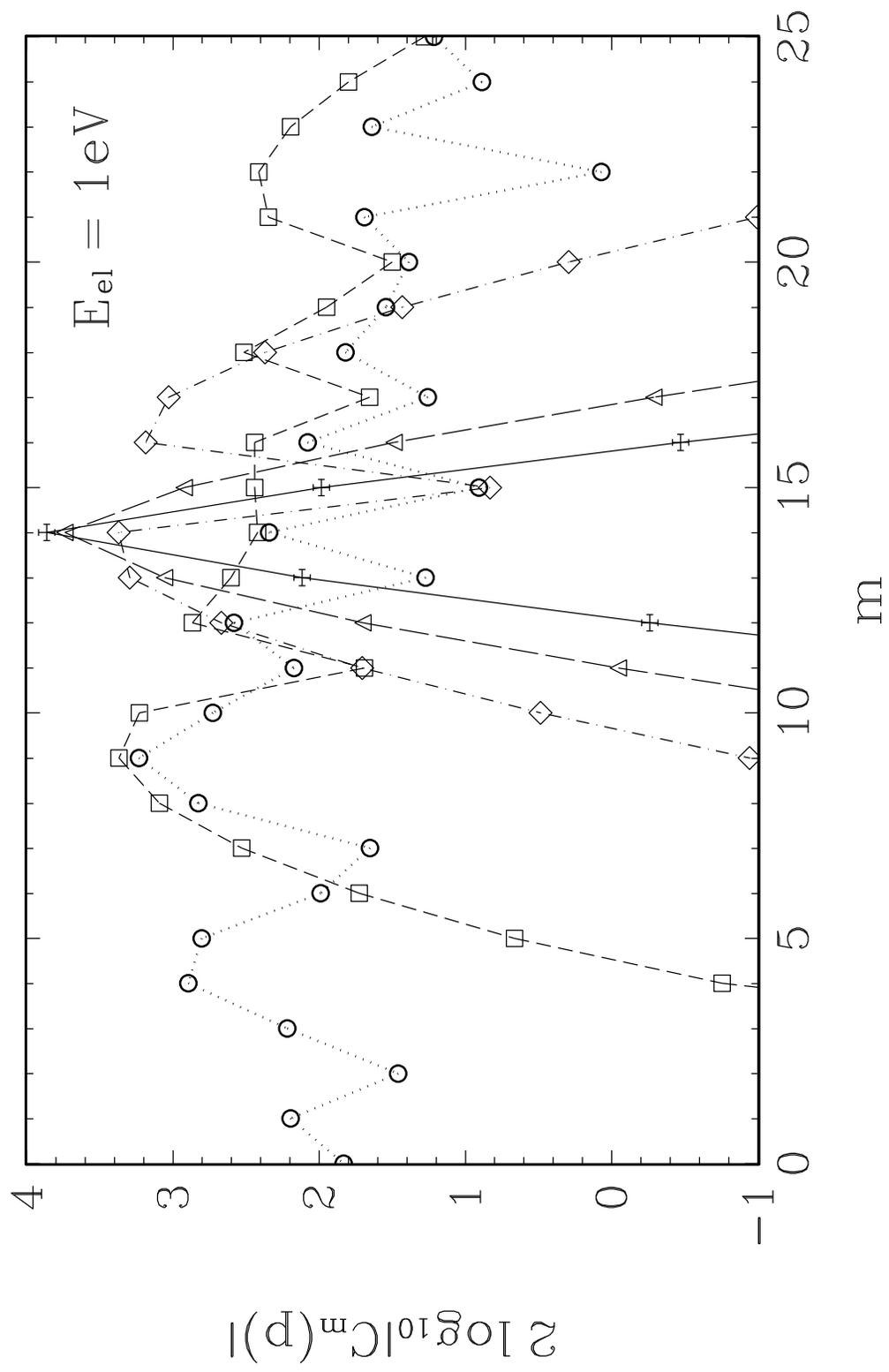
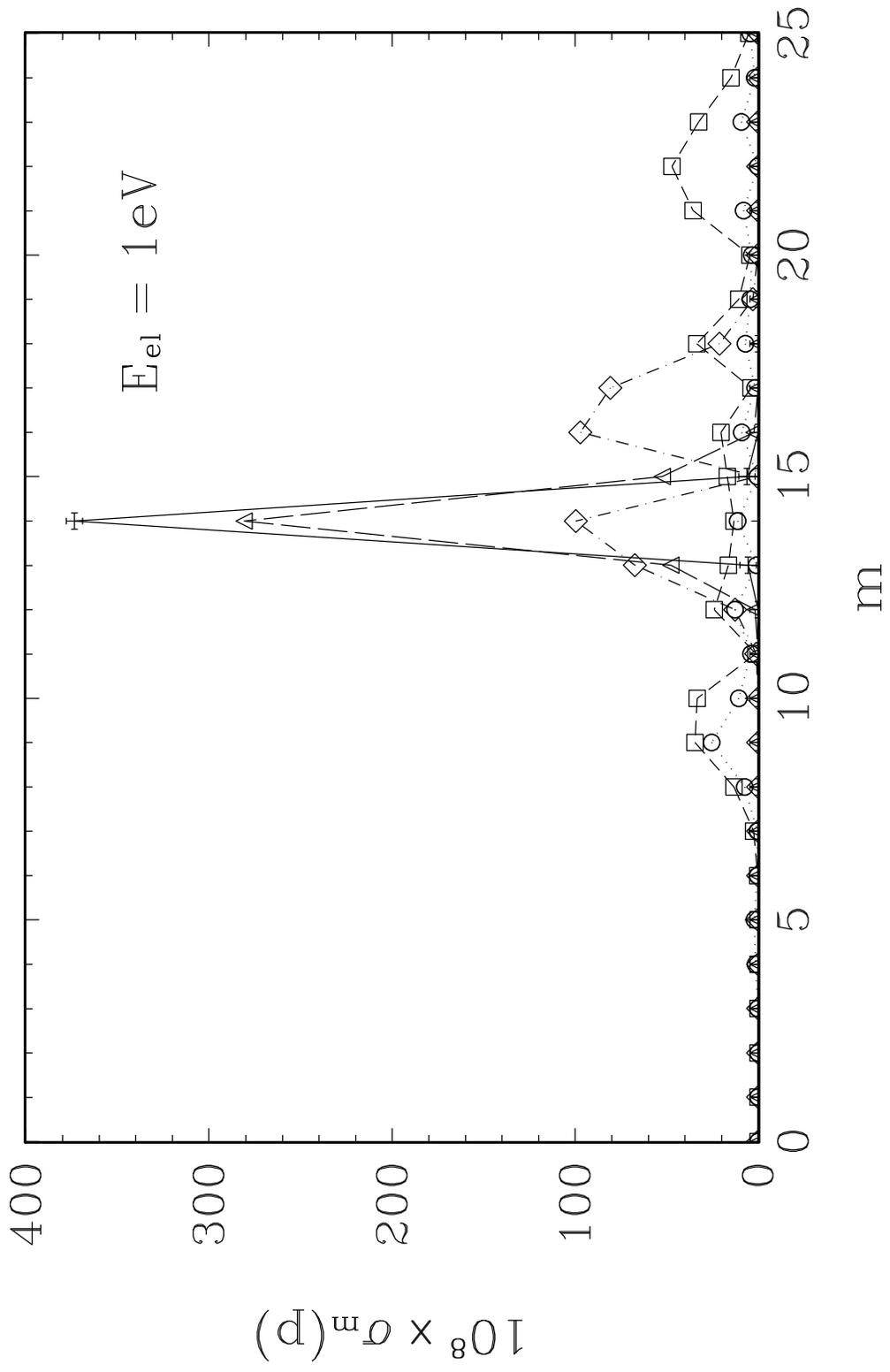
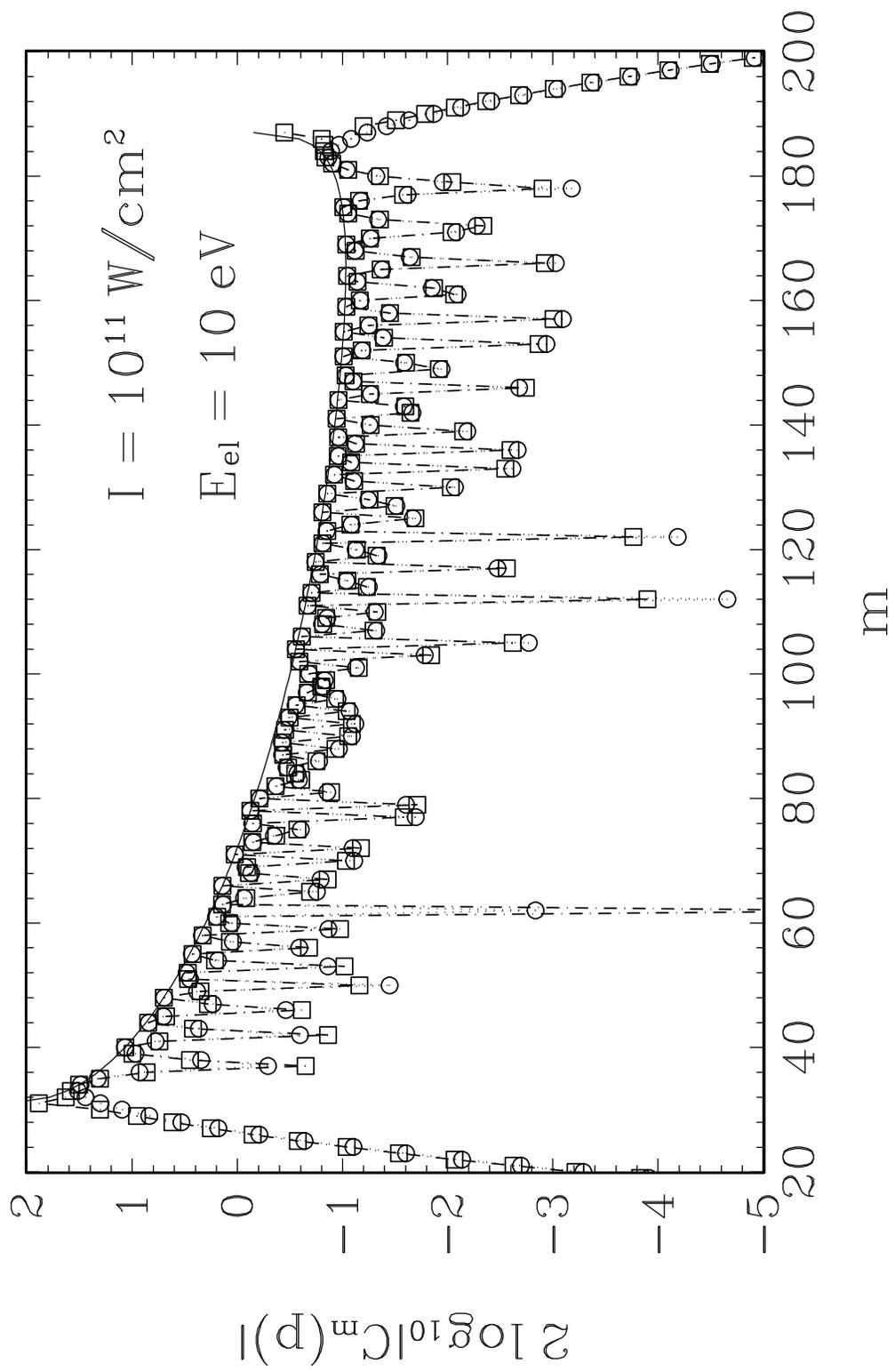
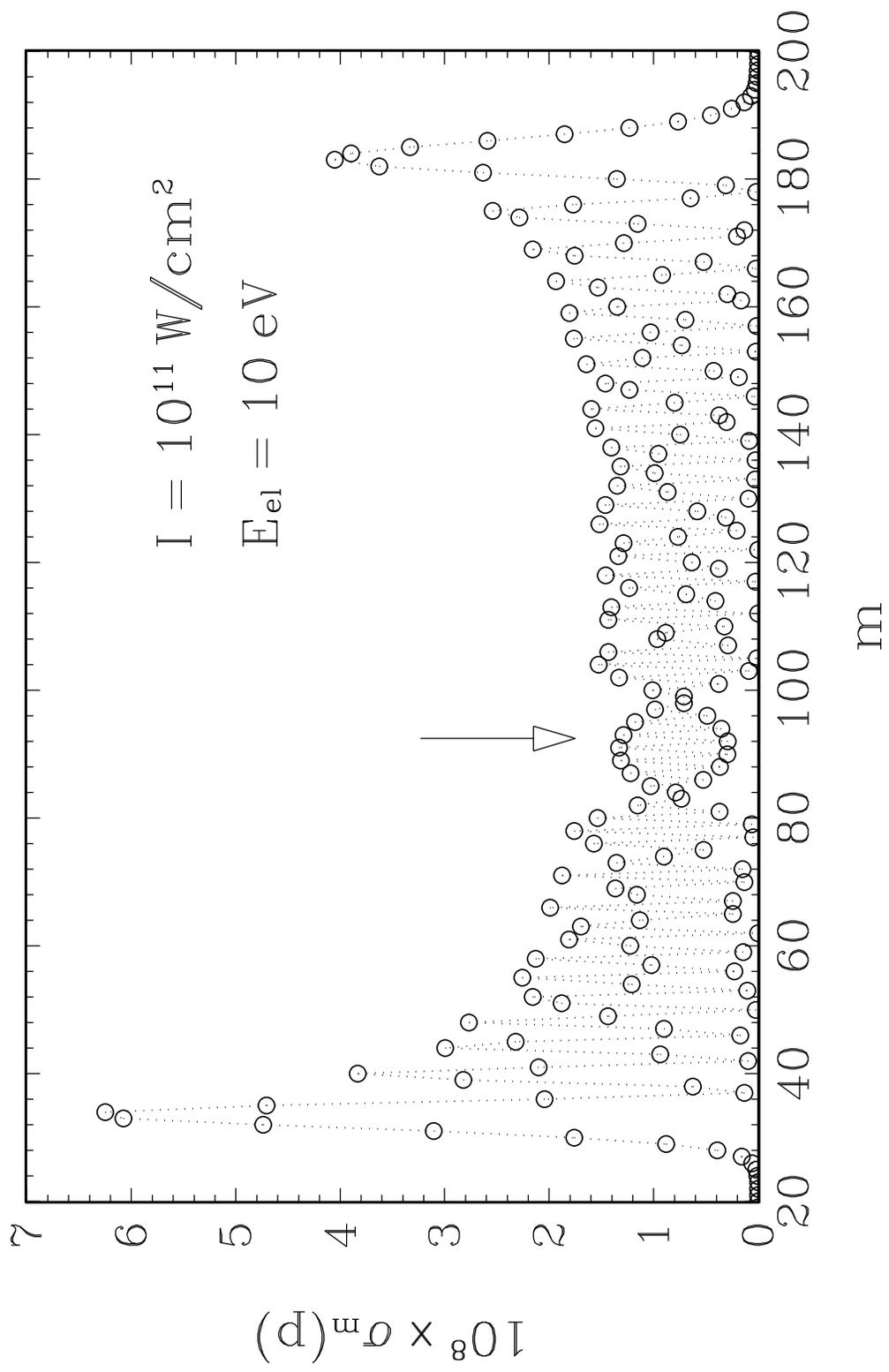
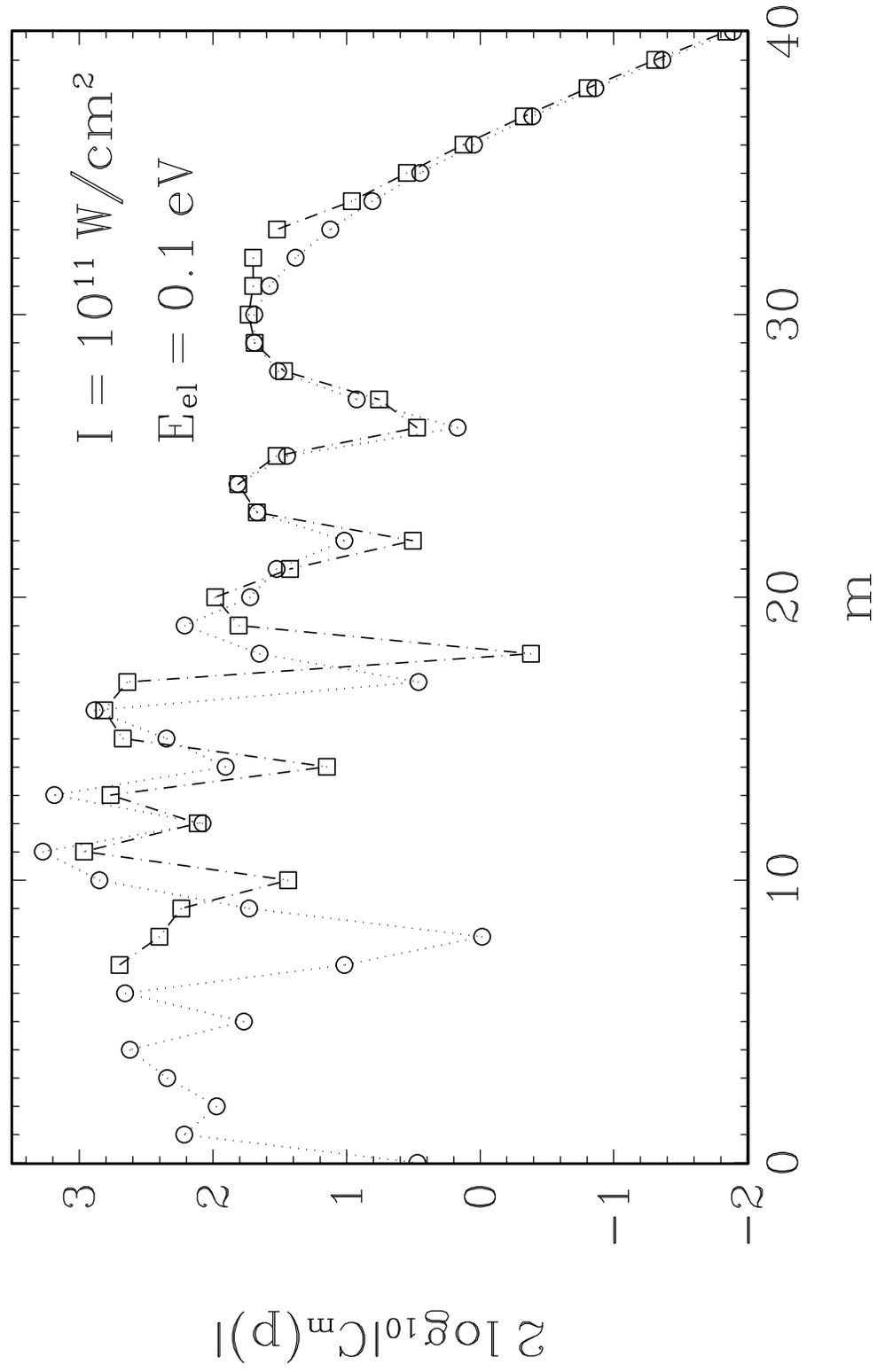
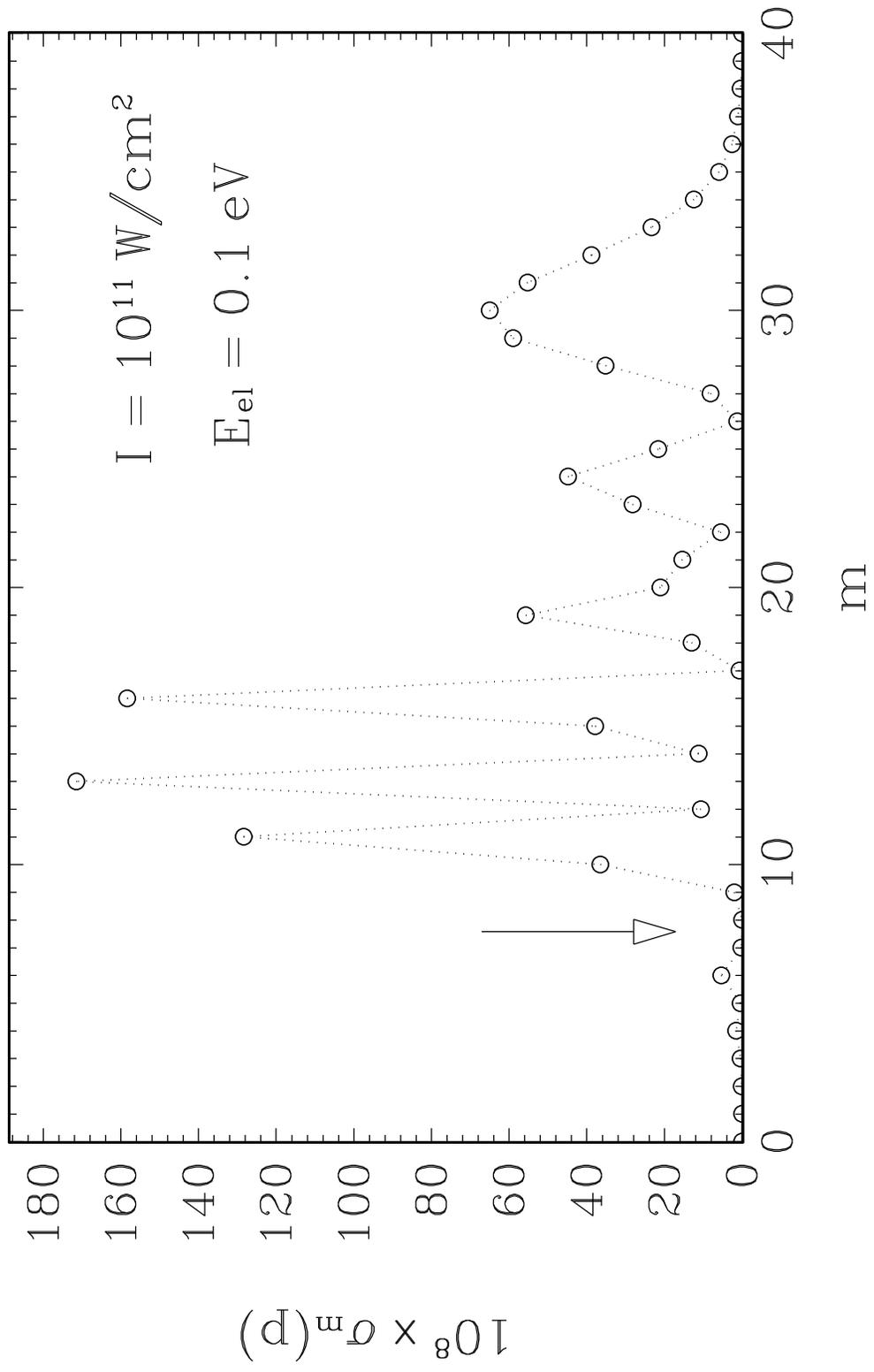
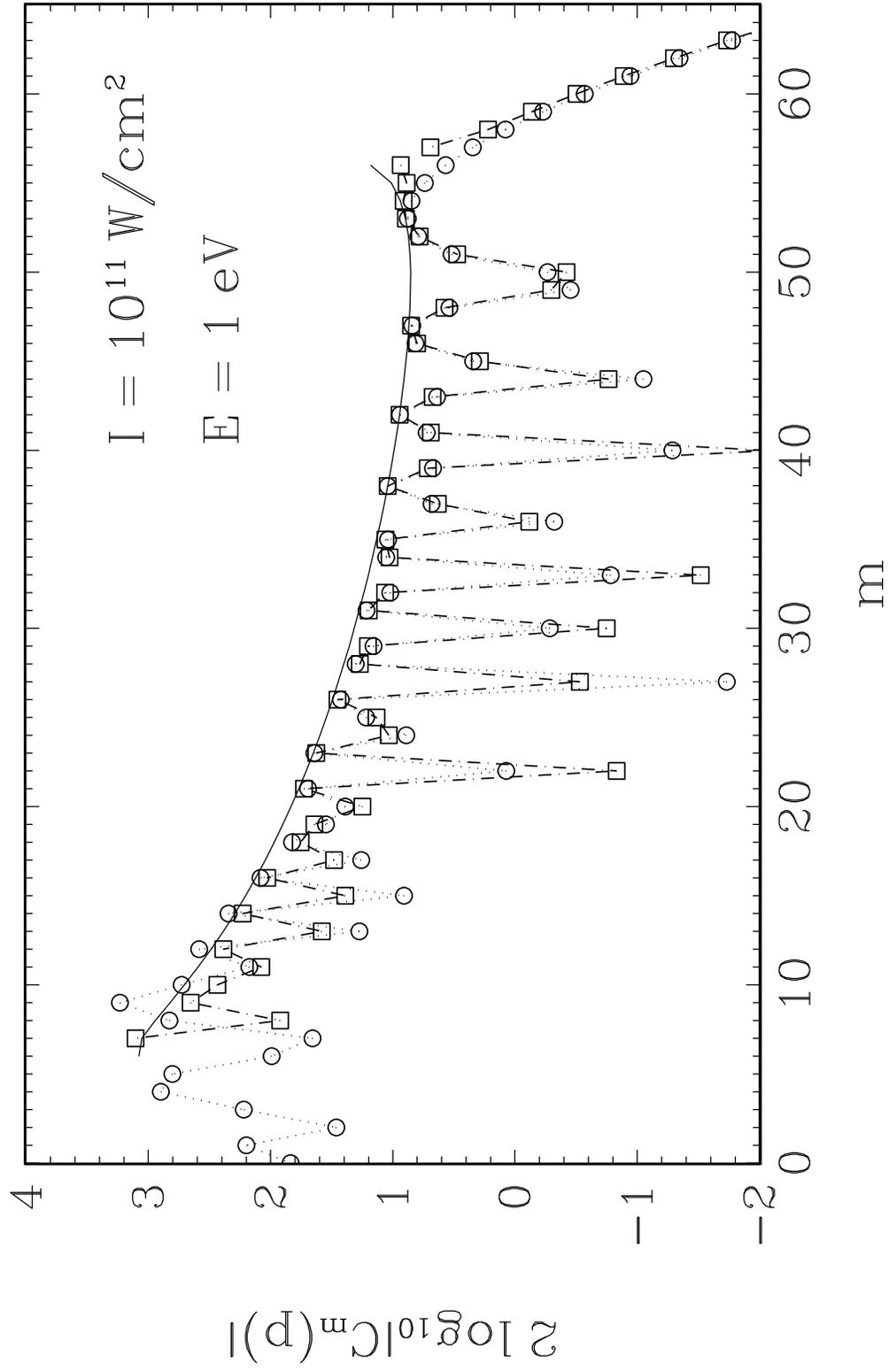
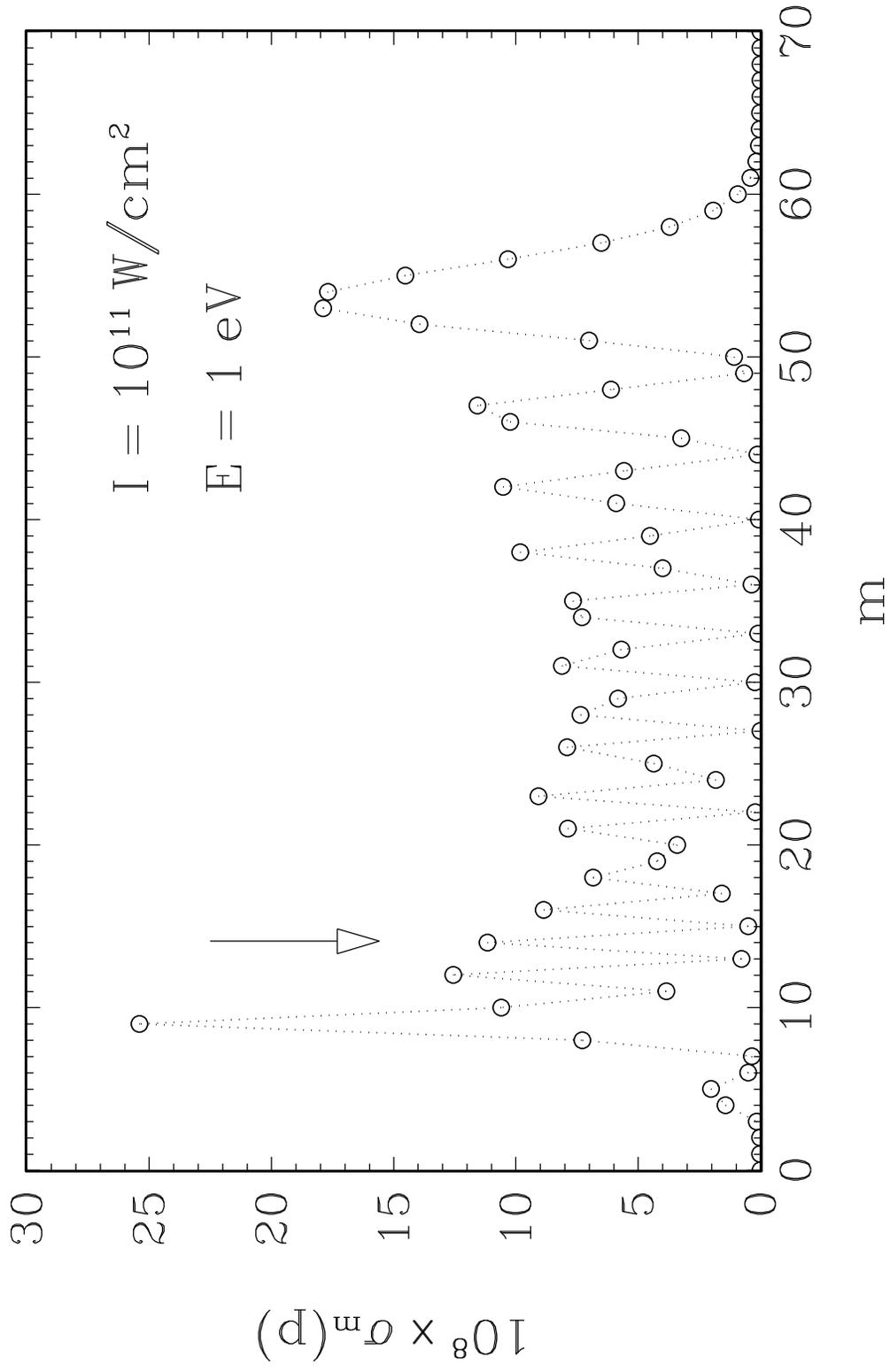
\end{document}